\documentclass[twocolumn,preprintnumbers,amsmath,amssymb,aps,nofootinbib]{revtex4-1}
\usepackage{graphicx}
\usepackage{physics}
\usepackage{color}
\usepackage{gensymb}
\usepackage{multibib}
\newcites{New}{References}

\begin{document}

\title{Fermion-Mediated Interactions Between Bosonic Atoms}
\author{B.J. DeSalvo}
\thanks{These authors contributed equally to this work.}
\author{Krutik Patel}
\thanks{These authors contributed equally to this work.}
\author{Geyue Cai}
\author{Cheng Chin}

\affiliation{James Franck Institute, Enrico Fermi Institute, and Department of Physics, \\ The University of Chicago, Chicago, IL 60637, USA}

\begin{abstract}

In high energy and condensed matter physics, particle exchange plays an essential role in the understanding of long-range interactions. For example, the exchange of massive bosons leads to the Yukawa potential \cite{Yukawa1935,Yukawa1937}. Phonon exchange between electrons gives rise to Cooper pairing in superconductors \cite{Tinkham2004}. When a Bose-Einstein condensate (BEC) of Cs is embedded in a degenerate Fermi gas of Li, we show that interspecies interactions can give rise to an effective trapping potential, damping, and attractive boson-boson interactions mediated by fermions. The latter, related to the Ruderman-Kittel-Kasuya-Yosida (RKKY) mechanism \cite{Ruderman1954}, results from a coherent three-body scattering process.  Such mediated interactions are expected to form novel magnetic phases \cite{De2014} and supersolids \cite{Buchler2003}.  We show that for suitable conditions, the mediated interactions can convert a stable BEC into a train of ``Bose-Fermi solitons"  \cite{Karpiuk2004,Santhanam2006}.
\end{abstract}

\maketitle

Interactions between cold neutral atoms are typically well-approximated by contact interactions and are characterized by a single parameter, the scattering length $a$.  Recent experiments utilizing highly magnetic atoms \cite{Lu2011}, Rydberg atoms \cite{Balewski2013}, and ground-state polar molecules \cite{Ni2008}, have stimulated great interest to probe novel quantum many-body states with long-range interactions. Examples include  quantum droplets \cite{Ferrier2016,Chomaz2016} and lattice spin phases \cite{Hazzard2014,Guardado2018,Zeiher2016}.

Another class of many-body systems that exhibit long-range interactions are quantum mixtures in which particle interactions are mediated by interspecies scattering.  Here, we consider the case of interactions between bosons mediated by a degenerate Fermi gas.  In the regime that the dynamics of the fermions are much faster than those of the bosons, an effective description for the bosons applies.  In this case, the mediated interactions are a spinless analog of RKKY interactions \cite{De2014,Suchet2017},
\begin{equation}
U(R) \propto - g_{BF}^2\frac{\sin R - R \cos R}{R^4},
\label{Eg1}
\end{equation}
where $R = 2 k_F R_B$, $R_B$ is the separation between bosons, $g_{BF} = 2 \pi \hbar^2 a_{BF} (\frac{1}{m_B} + \frac{1}{m_F})$ is the interspecies interaction strength, $a_{BF}$ is the interspecies scattering length, $m_{B(F)}$ is the boson (fermion) mass, $\hbar$ is the reduced Planck's constant, and $k_F$ is the Fermi wavevector. The interaction is attractive at short distance regardless of the sign of $g_{BF}$, and is oscillatory at long-range with the length scale of $\pi/k_F$, which corresponds to 1 $\mu m$ in our experiment.  

\begin{figure}[h]
\includegraphics[clip,trim = 0in 0.0in 0.0in 0.0in,width=3.4 in]{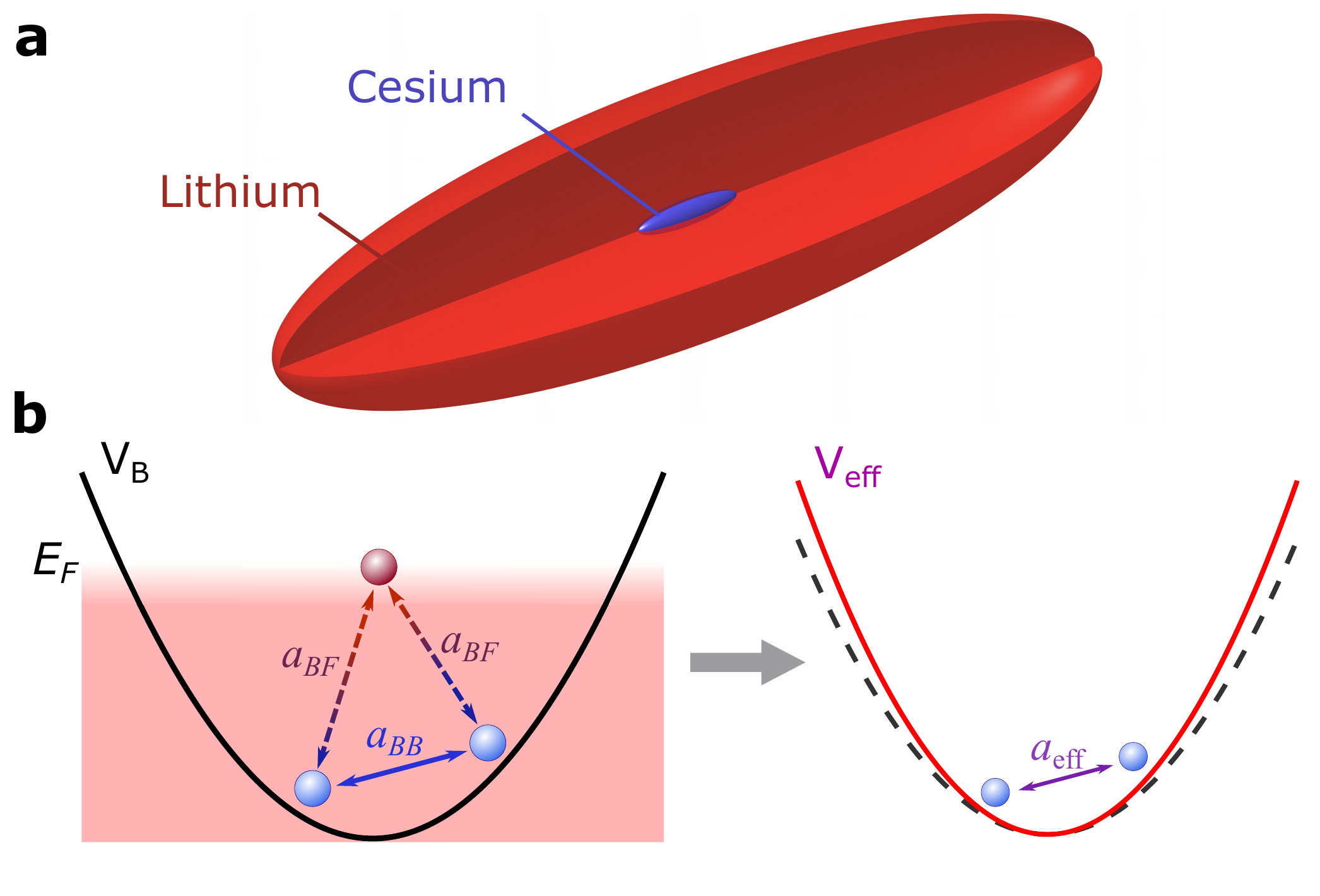}
\caption{Mediated interactions between Cs atoms by exchanging Li atoms near the Fermi surface. \textbf{a} A cigar-shaped BEC of heavy Cs bosons (blue) is fully immersed in a degenerate Fermi gas of light Li atoms (red). \textbf{b} Two Cs atoms (blue balls) can interact directly with one another with scattering length $a_{BB}$ or via secondary interactions with a Li atom (red ball) near the Fermi surface with energy $E_F$. By eliminating the fermionic degrees of freedom, the bosons can be described by an effective trapping potential $V_{\text{eff}}$ and an effective scattering length $a_{\text{eff}}$ (see text).}
\label{Fig1}
\end{figure}

In this work, we demonstrate the effect of boson-boson interactions mediated by fermions in a degenerate Fermi gas.  By using a mixture of fermionic $^6$Li and bosonic $^{133}$Cs, our system offers the tunability needed to observe the effect of such interactions, ultimately leading to the formation of a Bose-Fermi soliton train. In our system, the combination of light fermions and heavy bosons with their associated large Fermi energy $E_F$ and small BEC chemical potential $\mu_B$ ensures a large separation of the relevant timescales ($\hbar/E_F \approx 15~\mu s$ for fermions and $\hbar/\mu_B \approx 500~\mu s$ for bosons). Furthermore, the large mass imbalance and difference in quantum statistics allows us to prepare a BEC that is fully immersed in the degenerate Fermi gas, as shown schematically in Fig.~\ref{Fig1}a.  For zero interspecies interactions, the density of the fermions within the BEC is nearly constant. For weak interactions, one may eliminate the fermionic degrees of freedom and obtain an effective energy density functional of the bosonic field $\Psi_B(r)$, given by \cite{Tsurumi2000, Chui2004,Santamore2008}
\begin{equation}
E = \frac{\hbar^2}{2m_B} |\nabla\psi_B(r)|^2+ V_{\text{eff}}(r)|\psi_B(r)|^2 + \frac{g_{\text{eff}}}{2}|\psi_B(r)|^4,
\label{Eq1}
\end{equation}
where $V_{\text{eff}}(r) = \sum_i \frac{1}{2}m_B \omega_{\text{eff},i}^2 r_i^2$ is the effective potential and $r_i = x,y,z$. The effective harmonic trapping frequencies $\omega_{\text{eff},i}$ and effective interaction strength $g_{\text{eff}} = 4\pi \hbar^2 a_{\text{eff}}/m_B$ satisfy the relations \cite{Tsurumi2000, Chui2004}
\begin{align}
\omega_{\text{eff},i}^2 &= \omega_{B,i}^2 -  \frac{3}{2}\frac{n_F}{E_F} \frac{m_F}{m_B} g_{BF} \omega_{F,i}^2 \label{Eq3}\\
g_{\text{eff}} &= g_{BB} - \xi \frac{3}{2}\frac{n_F}{E_F} g_{BF}^2, \label{Eq4}
\end{align}
where $\omega_{B(F),i}$ denote the bare harmonic trapping frequencies of the bosons (fermions) and $g_{BB} = 4 \pi \hbar^2 a_{BB}/m_B$ is the boson-boson interaction strength for scattering length $a_{BB}$.  We note that owing to the large mass imbalance and large Fermi energy in our system, Eq. (\ref{Eq3}) is well approximated by a linear dependence of $\omega_{\text{eff}}$ on $a_{BF}$ for all scattering lengths probed in our experiment.   

The dimensionless constant $\xi$ in Eq. (\ref{Eq4}) characterizes the strength of the fermion-mediated interaction.  Calculations based on hydrodynamics \cite{Tsurumi2000}, path integrals \cite{Chui2004}, and diagrammatic expansion \cite{Santamore2008} yield $\xi = 1$, where as a two-body scattering model predicts $\xi = \pi^3$ \cite{De2014}. 

From Eqs. (2-4), it is apparent that the presence of the Li degenerate Fermi gas is expected to alter the dynamics of the Cs BEC in two important ways.  The BEC will experience both a modified  harmonic trapping force as well as a modified scattering length $a_{\text{eff}}$.  The former effect can be understood in the mean-field picture.  As the degenerate Fermi gas is also trapped in the harmonic potential, the spatially dependent density yields a mean-field potential on the BEC.  For attractive (repulsive) interactions, the BEC experiences a stronger (weaker) harmonic confinement.

The mediated interaction, described by the second term of Eq.~(\ref{Eq4}), is a genuine three-body scattering effect with an energy $E\propto -g_{BF}^2 n_B^2 n_F$. A microscopic picture giving rise to these mediated interactions is sketched in Fig.~\ref{Fig1}b.  They arise when two bosons exchange a fermion near the Fermi surface via the usual two-body s-wave scattering (dashed lines).   We note here that the mediated interactions are always attractive regardless of the sign of $g_{BF}$. This can be understood in the mean-field picture as follows: For repulsive interspecies interactions, fermions are repelled from the BEC, and thus the BEC feels an additional attraction towards its center where the fermion density is the lowest.  On the other hand, for attractive interspecies interactions, fermions are attracted to the center of the BEC, and as a result the BEC experiences an additional attraction towards its center where the fermion density is the highest.

\begin{figure}[h!]
\includegraphics[clip,trim = 0.0in 0.0in 0.0in 0.0in,width=3.4 in]{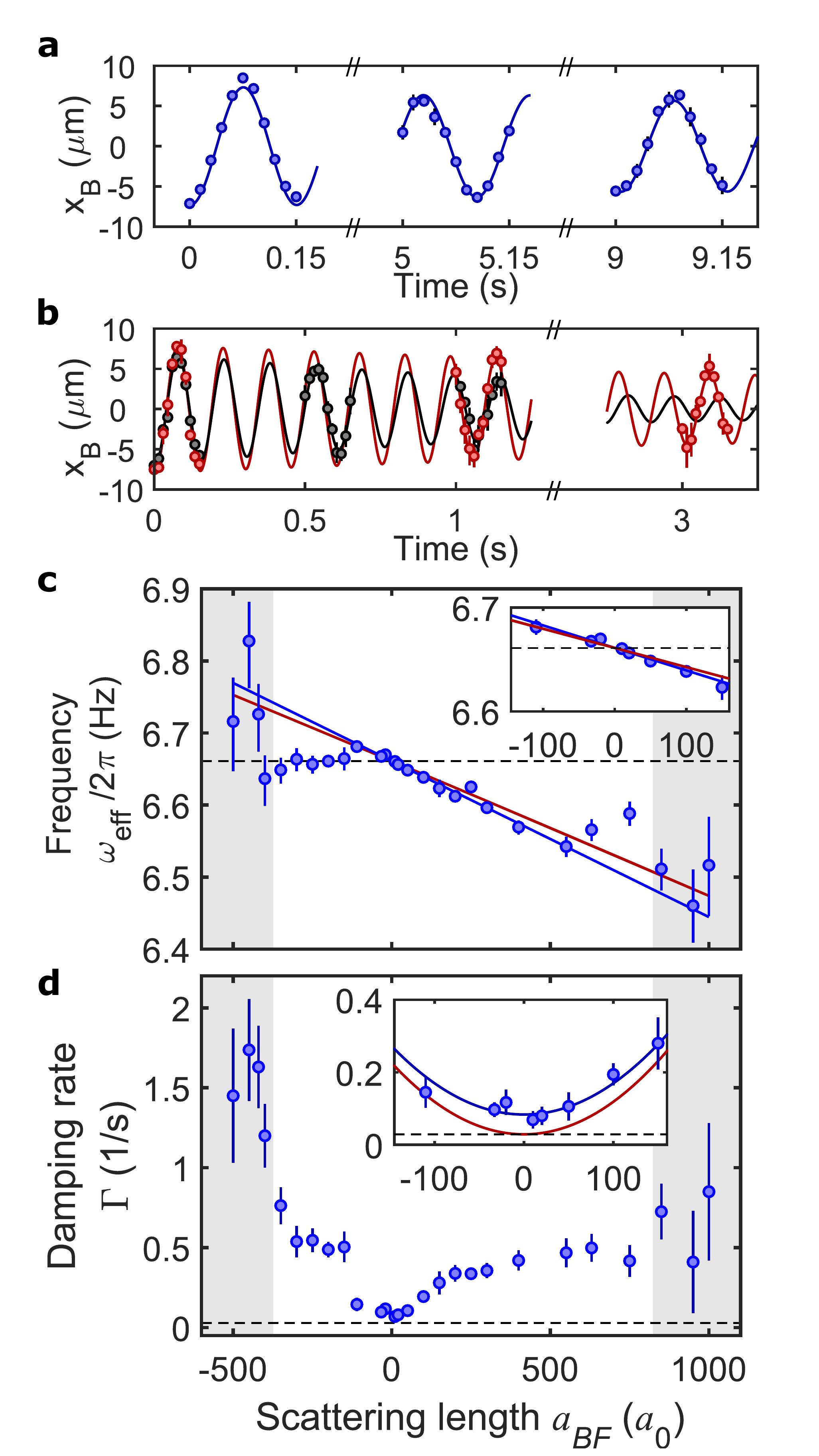}
\caption{Dipole oscillations of a Cs BEC immersed in a Li degenerate Fermi gas.  \textbf{a} Without Li present, we observe long-lived dipole oscillations along the weakly confined $x-$axis.  The data is well described by damped harmonic oscillation with very low damping $\Gamma<1/30~s$ (blue line).  \textbf{b} When the BEC is embedded in the Fermi gas, a shift in the oscillation frequency as well as an increase of the damping are observed. Example traces are shown for $a_{BF} = 100~a_0$ (red) and   $a_{BF} = 630~a_0$ (black). \textbf{c} Oscillation frequency and \textbf{d} damping rate of the BEC. For small $\abs{a_{BF}}$, the data is well fit by linear and quadratic curves (blue lines) and are consistent with the predictions (red lines), see Eq.~(3) and Methods. Shaded areas show the magnetic field ranges with significant heating where the BEC is heated into a normal gas. Frequency and damping of a bare Cs BEC (dashed lines) serve as references for comparison.}
\label{Fig2}
\end{figure}

To observe the effective trapping frequency as well as the mediated interactions, we perform measurements at different scattering lengths $a_{BB}$ and $a_{BF}$ near intra- and interspecies Feshbach resonances \cite{Berninger2013,Tung2013} (see Methods).  In our system, following the procedure outlined in Refs.~ \cite{Johansen2017, DeSalvo2017}, we prepare quantum degenerate mixtures of $2 \times 10^4$ Li and $3 \times 10^4$ Cs atoms in their absolute ground state at magnetic field $B \approx 900$~G. Both species are radially trapped in a single laser beam and weakly confined magnetically in the $x-$direction.   

To measure the effective harmonic trapping frequency, we excite dipole oscillations of the Cs BEC along the weakly trapped axis \cite{Ferlaino2003,Ferrier-Barbut2014,Roy2017}. In the absence of Li, we observe long-lived oscillations with a  decay time of $\approx 30$ s (see Fig.~2a). This low background damping rate allows us to precisely determine the trapping frequency of the Cs (see Methods).  

In the presence of Li, we observe small shifts in the oscillation frequency $\omega_{\text{eff}}$ and enhanced damping $\Gamma$ as $a_{BF}$ deviates from zero (see  Fig.~\ref{Fig2}b). In the range of $-100~a_0 < a_{BF} < 500~a_0$, we observe a linear dependence of the frequency of the oscillation on the scattering length $a_{BF}$ (see Fig.~\ref{Fig2}c). A fit to the data in this range shows a negligible shift $<~0.1\%$ at $a_{BF} = 0$ and a slope of $-~0.22(2)$ mHz/$a_0$ in fair agreement with prediction of $-0.18$ mHz/$a_0$ from Eq.~(\ref{Eq3}) . Outside this range, a non-linear behavior develops. The frequency shifts first reduce toward the bare Cs frequency and then, for very large $|a_{BF}|$, the BEC is heated to a normal gas due to recombination loss (see shaded area in Figs.~2c and d), and displays even larger shifts. Such non-linear behavior at large $a_{BF}$ shows the breakdown of the mean-field theory and demands further theoretical investigation. 

The damping of dipole oscillations shows interesting behavior as well. For small scattering lengths, collisions between the BEC and the Fermi gas are scarce, and yield a weak friction that damps the oscillation.  The damping is expected to be proportional to the collisional cross-section $\propto a_{BF}^2$ \cite{Ferrari2002,Lous2018}.  In the range of $-100~a_0 < a_{BF} < 100~a_0$, our data agree well with the prediction $\Gamma = \kappa a_{BF}^2$ (see inset of Fig. \ref{Fig2}d and Methods). A fit to the data in this range yields a curvature of $\kappa=8.5(2.1)\times 10^{-6} /s\cdot a_0^2$, which excellently agrees with the prediction of $\kappa = 9 \times 10^{-6} /s\cdot a_0^2 $. At large scattering lengths, the damping appears to saturate and then increases again when the sample is heated to a normal gas. While a more complete model is needed to describe the motion of a BEC strongly interacting with a degenerate Fermi gas, we note that for small scattering lengths ($|a_{BF}| < 100~a_0$), the mean-field theory offers very good quantitative predictions. 
 
To measure the mediated interactions, we first characterize the bare scattering length of the Cs atoms based on the equilibrium Thomas-Fermi radius $R_{TF}\propto \frac{1}{\omega_x}(N_0 a_{BB})^{1/5}$ in the weakly confined axis, where $N_0$ is the number of atoms in the BEC. In the relevant range of magnetic field the scattering length is approximately linear with a zero crossing at $B=880.29(8)$~G, as shown in Fig. 3a.  Our result is in good agreement with the calculation in Ref.~\cite{Berninger2013}.

\begin{figure}[h]
\includegraphics[clip,trim = 0in 0in 0in 0in,width=3.4 in]{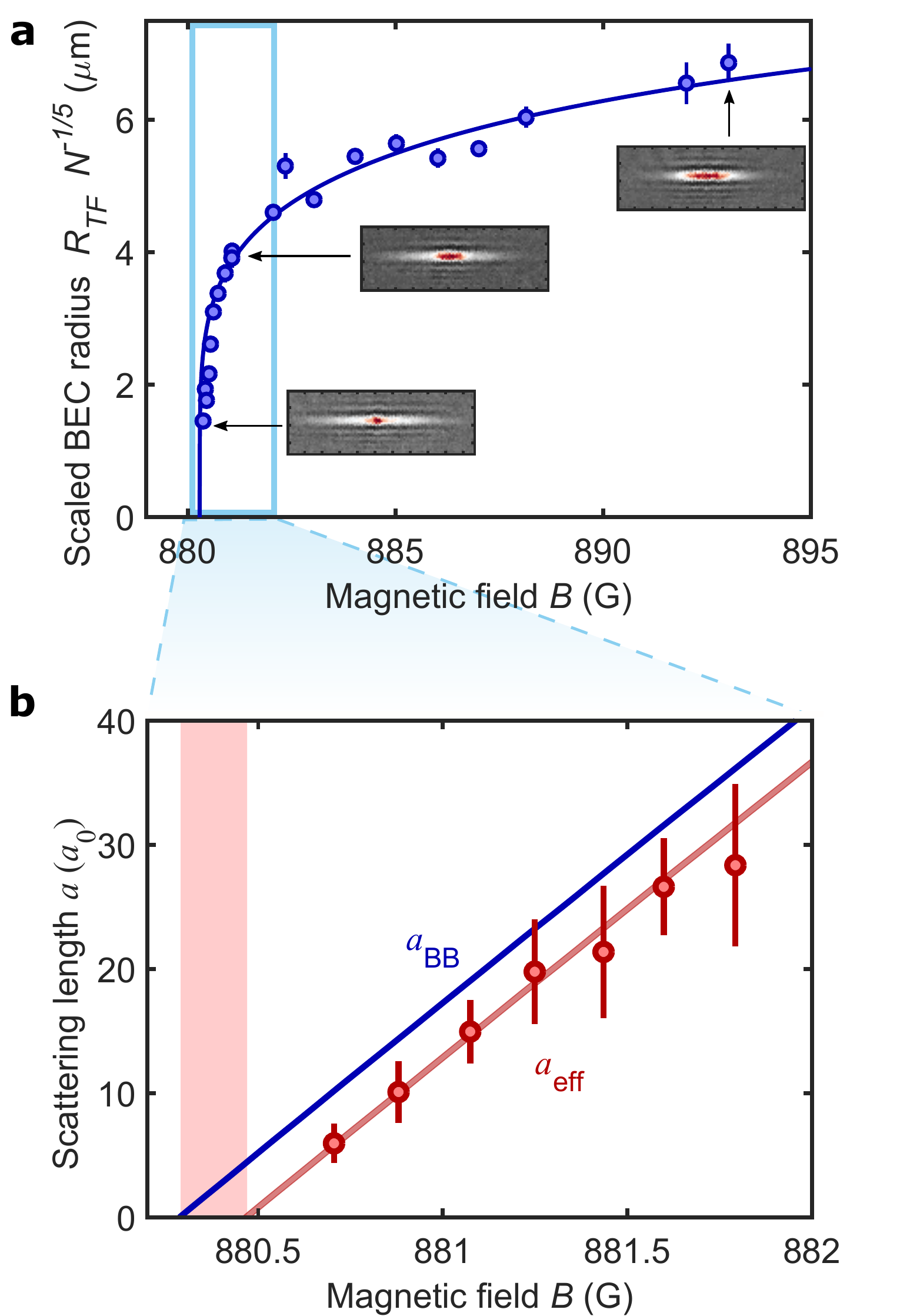}
\caption{Bare and effective Cs-Cs scattering length.  \textbf{a} We measure the \textit{in situ} Thomas-Fermi radius $R_{TF}$ of a Cs BEC in the weakly confined direction.  Without Li in the trap (blue circles), we find good agreement with theory when scaled appropriately to the number of atoms in the BEC (blue line).  Characteristic images from the measurement are shown as insets.  \textbf{b}  At small $a_{BB}$, we perform a relative measurement of $R_{TF}$ with and without Li present to extract the effective scattering length $a_{eff}$ (red circles).   Our result is consistent with a small negative offset from $a_{BB}$ (blue line) indicating attractive mediated interactions.  A fit to the data (red line) yields fair agreement with theory (see text).}
\label{Fig3}
\end{figure}

Armed with the calibrated scattering length, we next determine the effective scattering length of the BEC embedded in the degenerate Fermi gas. Since the mediated interactions are weak, we perform the measurement in the range of $|a_{BB}|\ll|a_{BF}|$ to enhance the effect.  We measure the change in the scattering length by directly comparing the BEC radius $R_{TF}$  with and without the degenerate Fermi gas present near the zero crossing $a_{BB}=0$ (see Methods).  We find that the size of the BEC is systematically smaller when Li is present, which is consistent with attractive mediated interactions (see Fig. 6). From our relative measurements, the effective scattering length $a_{\text{eff}}$ displays a negative and constant offset from the bare Cs scattering length throughout the measured range (see Fig.~ \ref{Fig3}b). The fitted offset is -4.4(3)(1.5) $a_0$, where the first and second uncertainties are statistical and systematic, respectively. Together with the known constant $a_{BF} = - 60~a_0$ \cite{Tung2013}, the measured offset yields a value of $\xi = 1.7(6)$ in Eq. (4), in fair agreement with the calculations in Refs.~\cite{Tsurumi2000, Chui2004,Santamore2008}.  The large deviation of our result from Ref. \cite{De2014} is likely because our experiment is not in the dilute limit assumed in the calculation.

These mediated interactions look minute, but can have a profound effect on the ground state of the Cs atoms. Labeled as the red shaded area in Fig.~\ref{Fig3}b, in particular, when $a_{BB}$ is small and positive, the mediated interactions can cause $a_{\text{eff}}$ to become negative. As an harmonically trapped BEC with attractive interactions can experience dynamic instability and collapse \cite{Dalfovo1999}, one expects that in this regime, the mediated interactions can render the BEC unstable as well. In our highly elongated trap, the BEC enters the quasi-1D regime for scattering lengths $a_{BB} < 4~a_0$, and Bose-Fermi soliton trains are predicted to form in this regime with sufficiently strong fermion-mediated attraction between bosons \cite{Karpiuk2004,Santhanam2006}. In our slow ramp experiments, we observe very large shot-to-shot fluctuations of the shape and size of the cloud in this region, indicating the onset of the instability.  

\begin{figure}
\includegraphics[clip,trim = 0.4in 1.1in 0.4in 0.65in,width=3.4 in]{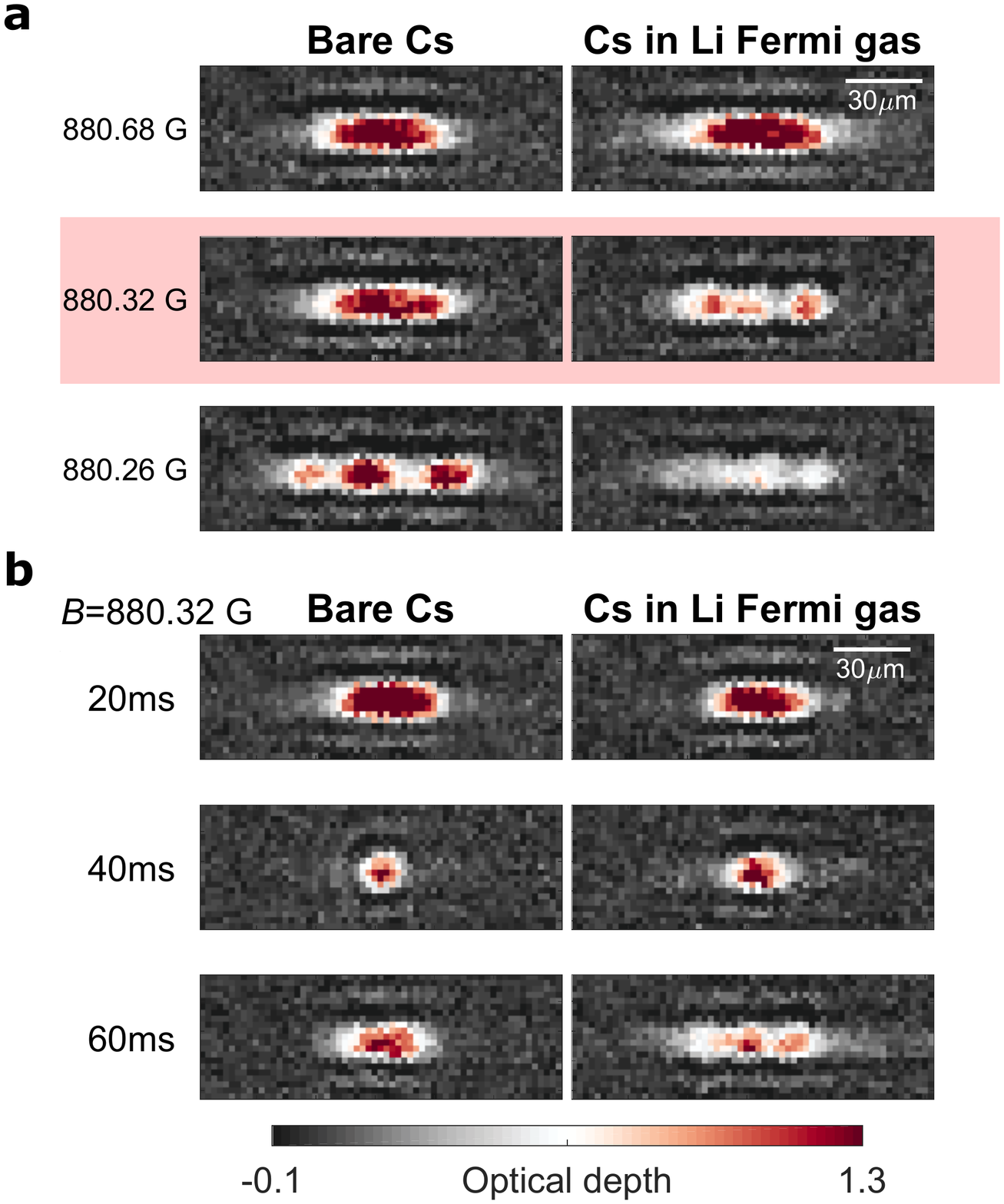}
\caption{Formation of Bose-Fermi solitons. \textbf{a} \textit{In situ} images of a Cs BEC 75~ms after a quench from $B = 885.5$~G ($a_{BB} = 120~a_0$) to lower fields indicated on the left. For positive $a_{\text{eff}}$, there is no qualitative difference when Li is added to the system.  When $a_{\text{eff}}$ becomes negative, the BEC collapses into a train of solitons with Li present, but remains stable without Li as shown for $B = 880.32$~G in the shaded red panels.  For $a_{BB}<0$, the BEC collapses with and without Li present. \textbf{b} Time sequence of the induced collapse at $B = 880.32$~G.  Without Li present, the BEC undergoes breathing oscillations.  With Li, the BEC first shrinks to a small size and afterwards breaks up into a train of solitons. }   
\label{Fig4}
\end{figure}

To probe the fermion-mediated instability, we perform a dynamic experiment by suddenly jumping the scattering length from a large positive value to values near the zero crossing of $a_{BB}$ (see Fig.~4 and Methods). \textit{In situ} images taken 75~ms after the quench show a qualitative difference for a bare Cs BEC and for a BEC embedded in the degenerate Fermi gas. In the absence of Li, we observe the typical formation of a soliton train for a final magnetic field below B $\approx$ 880.26~G, where $a_{BB} < 0$.

In the presence of Li, we observe a collapse below B $\approx$ 880.34~G, where $a_{BB} > 0$. This observation is consistent with the results shown in Fig. \ref{Fig3} that at this field the effective scattering length is negative $a_{\text{eff}}<0$ and thus the stability of the BEC has been compromised. The observed soliton train is likely comprised of correlated bosonic and fermionic density waves \cite{Karpiuk2004,Santhanam2006}. However, we could not see a clear density modulation of the Fermi gas due to limited signal-to-noise of our measurement.

The induced collapse in this interesting regime is further supported by the time evolution of the BEC following a quench. After preparing the sample at $B = 885.5$~G where $a_{BB} = 120~a_0$, we abruptly change the field to 880.32~G, where the effective scattering length is expected to be negative $a_{\text{eff}}<0$, and monitor the subsequent dynamics of the BEC. Without the Fermi gas $a_{BB}>0$ the BEC undergoes breathing oscillations without qualitative changes in the structure of the cloud.  With Li present, the BEC first contracts as if starting a breathing oscillation, however, upon expansion, the BEC is fractured into a train of of 3 to 4 solitons.   This observation clearly indicates that the fermion-mediated interactions can destablize a weakly-interacting BEC. In other words, the ground state of the bosons can be altered by the fermion-mediated interactions. 

While the fermion-mediated interactions amount to only an effective change of the scattering length by -4 $a_0$ in our experiment, the scaling of the mediated interaction $a_{BF}^2n_B^2n_F$ suggests strong influence on systems near an interspecies Feshbach resonance as well as in systems with high local density, for example, in an optical lattice. This promises exciting future work exploring the long-range nature of such interactions to ultimately probe novel quantum phases beyond contact interactions.  

\textbf{Acknowledgements}  We thank M. Gajda for useful discussions and M. McDonald for a careful reading of the manuscript. This work was supported by NSF Grant No. PHY-1511696 and the University of Chicago Materials Research Science and Engineering Center, which is funded by the National Science Foundation under Grant No. DMR-1420709.

\section*{Methods}
\subsection*{Dipole Oscillation Experiments}
For the frequency and damping measurements presented in Fig. 2, we begin our experiment by preparing a quantum degenerate mixture of $2\times 10^4$ Li and $3\times 10^4$ Cs at rest in a harmonic trap with trapping frequencies of $\omega_{B,i}/2\pi$ = (6.65, 100, 140) Hz for Cs and $\omega_{F,i}/2\pi$ = (34, 320, 320) Hz for Li.  The sample is prepared near the interspecies Feshbach resonance at either $B =  892.3$ G or $B = 893.2$ G with $a_{BF} = 300~a_0$ or $ -300~a_0$ , respectively (see Fig. 5). We then apply a magnetic field gradient of 26 mG/cm along the $x$-axis of our trap which slightly displaces the center of both clouds by about $5~\mu m$. The magnetic field gradient is suddenly removed which starts a dipole oscillation of both species. Due to the small displacement, the relative velocity of the center of mass of Li and Cs remains small compared to the speed of sound of the Cs BEC.  
	
After 200 ms, the Li oscillation damps to near zero at which time we rapidly jump the magnetic field to the final target value. After a variable hold time, we  measure the center of mass position of the Cs BEC after a time-of-flight expansion. The center  position as a function of hold time is fit to a damped sinusoidal wave to extract the damping rate and frequency.

\subsection*{Coupled Oscillator Model}
To understand the damping rate of the oscillation measurements in the limit of small interspecies scattering length $a_{BF}$, we apply the  coupled oscillator model described in Ref. \cite{Ferrari2002}. In this model, the center of mass position $x_B$ of the BEC is described by \[\ddot{x}_B = -\omega^2 x_B  - \frac{4}{3}\frac{m_FN_F}{(m_F+m_B)(N_F+N_0)}\Gamma_{\text{coll}}\dot{x}_B,\] where $N_F$ is the number of fermions and $\Gamma_{\text{coll}}$ is the collision rate. The collision rate is given by \[\Gamma_{\text{coll}} = \sigma v \left( \frac{1}{N_F}+\frac{1}{N_0}\right)\int n_B n_Fd^3x,  \] where $n_{F(B)}$ is the density of the fermions (bosons), $\sigma = 4 \pi a_{BF}^2$ is the collision cross section, and the relative velocity $v$ is approximately given by the Fermi velocity, $v \approx v_F = \hbar k_F/m_F$. For a BEC immersed in a degenerate Fermi gas, the damping rate is approximately \[ \Gamma = \kappa a^2_{BF},\] where $\kappa= \frac{8\pi}{3} \frac{m_F}{m_F + m_B} v_F n_F$.

\subsection*{\textit{In situ} Size Measurements}
To measure the size of the Cs BEC we perform \textit{in situ} absorption imaging. We first apply a $20 \,\,\mu s$ resonant microwave pulse to transfer some population from the $\ket{F,m} =\ket{3,3}$ ground state to $ \ket{4,4}$, where $F$ and $m$ denote the total angular momentum and the magnetic quantum number, respectively. We then apply a $100\,\,\mu s$ imaging pulse resonant with the $\ket{4,4} \rightarrow \ket{5,5}$ transition.  

We use a short microwave pulse to keep the optical density low and minimize imaging artifacts, as well as to prevent the sample from spending too long in the $\ket{4,4}$ state, where strong intraspecies repulsion \cite{Leo2000} increases the cloud size.

\begin{figure}
\includegraphics[width=3.4in,trim={0.6in 10cm 2.5cm 10cm},clip]{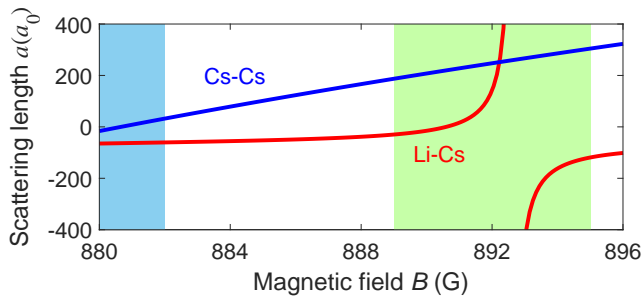}
\caption{Cs-Cs scattering length $a_{BB}$ (blue line) and Li-Cs scattering length $a_{BF}$ (red line) over the magnetic field range studied in this work. The interspecies Feshbach resonance near $B=892$~G is used for sample preparation and the effective trapping frequency measurements, indicated by the light green shaded area. The Cs-Cs zero crossing near $B = 880$~G is used for the effective scattering length measurements, indicated by the blue shaded area.  In this region, the Li-Cs scattering length is nearly constant $a_{BF} = -60~a_0$ \cite{Tung2013}.}
\label{scatlength}
\end{figure}

\subsection*{Scattering Length Characterization}
To characterize the bare scattering length of a pure Cs BEC, we evaporatively cool Cs at $B=893.5$ G, and after forming a BEC we adiabatically ramp the magnetic field to a target value over 1.4 s. Finally, we hold for 300 ms before taking an \textit{in situ} absorption image. To extract the BEC width from the \textit{in situ} image, we fit a line-cut through the center of the cloud to a bimodal density profile with a Thomas-Fermi distribution for the BEC and a Gaussian distribution for the thermal fraction.

Guided by the model presented in  Ref.~\cite{Berninger2013} for the value of the Cs scattering length, we fit our width measurements as a function of magnetic field assuming that the scattering length is approximately linear through the zero crossing. The data in Fig. 3a was obtained from two experimental runs, and to account for drift in our experiment we perform a joint fit of the two data sets with  different atom number calibrations for each set. Additionally, we omit data taken at very small scattering lengths (below $B = 880.60$ G), where we find departures from the Thomas-Fermi approximation due to the limitations of our imaging resolution and possible effects of the quasi-1D geometry.

\begin{figure}
\includegraphics[width=3.4in, trim={4cm 8cm 5cm 9cm},clip]{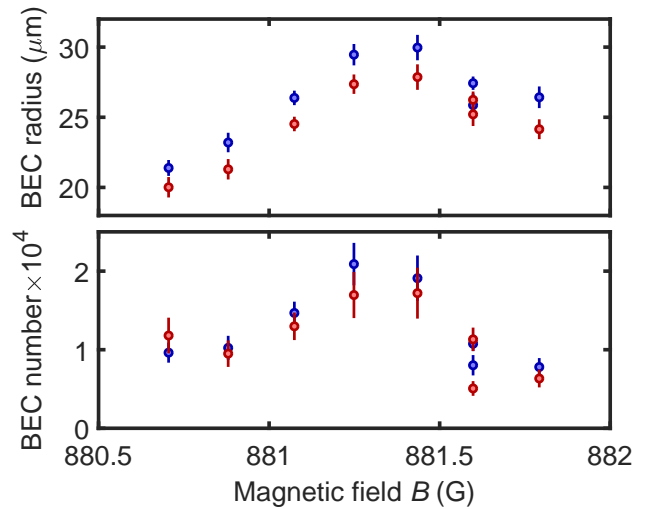}
\caption{Raw data for the scattering length difference measurement. Top: Measured  radius of the Cs BEC with (red circles) and without (blue circles) Li. Bottom: Measured BEC number with (red circles) and without (blue circles) Li. At every field, the size of the BEC with Li present is smaller than the corresponding measurement without Li. However, overall experimental drift makes absolute comparison difficult. Calibrating each measurement point-by-point as described in the text allows us to extract the difference in scattering length cleanly, as can be seen in Fig. 3b.}
\label{rawdata}
\end{figure}

\subsection*{Scattering Length Difference Measurement}
For the scattering length shift measurements presented in Fig. 3b, the mixture is first prepared at a magnetic field of $B=892.1$ G and ramped over 50 ms to $B=890.6$ G where
$a_{BF} = 0$ (see Fig. 5). At this field, we either remove Li atoms with a 100 $\mu s$ resonant light pulse or leave them in place. This procedure ensures the number of Cs atoms in the BEC is similar with or without Li present. The rest of the experiment is identical to the experiment described in the previous section.

The BEC number and raw width measurements associated with Fig. 3b are shown in Fig. \ref{rawdata}.  The observed BEC sizes are clearly smaller when Li is present for each magnetic field, which confirms the negative contribution from the mediated interactions. The decrease in cloud size at larger fields is due to atom number variation. 

 Using the Thomas-Fermi approximation and the model in Ref. \cite{Berninger2013} with a zero-crossing chosen according to our experimental determination, the BEC number can be calibrated as \cite{Dalfovo1999}

\[N = \frac{a_{\text{ho}}}{15a_{BB}}\left(\frac{m_B\omega_x^2R_x^2}{\hbar\bar{\omega}}\right)^{5/2}, \]
where $a_{\text{ho}} = \sqrt{\frac{\hbar}{m\bar{\omega}}}$ is the mean harmonic  oscillator length, $\omega_x / 2\pi = 6.65 $ Hz is the measured long-axis trap frequency, and $R_{TF}$ is the long-axis Thomas-Fermi radius. The transverse trap frequencies are $\omega_y/2\pi=130$ Hz and $\omega_z/2\pi=150 $ Hz, and $\bar{\omega} = (\omega_x\omega_y\omega_z)^{1/3}$ is the geometric mean of the trap frequencies.

Using the number calibration from the \textit{in situ} size without Li present, we can then accurately measure $N_0$ when Li is present from only the time of flight images. This allows us to extract the shift in scattering length from their \textit{in situ} size.

\subsection*{Soliton Experiments}

For the experiments shown in Fig 4, we begin the experiment by preparing our mixture at $B=892.1$ G. Then,  we ramp the field to $B=890.6$ G ($a_{BF}=0$) and either remove the Li atoms or leave them in place. Afterwards, we adiabatically ramp the field over 1.4 s to $B=885.5$ G where the Cs-Cs scattering length $a_{BB} = 120~a_0$.  We then suddenly jump the magnetic field to a target value. The magnetic field settles in 5 ms, which is fast compared to the trap oscillation period $2\pi / \omega_x = 150$ ms. 

For the images shown in Fig. 4a, we perform the experiment for the  target fields indicated in each panel, then allow the atoms to evolve for 75 ms before \textit{in situ} imaging the Cs BEC. For the images shown in Fig 4b, we instead choose one target field and hold for the duration indicated in each panel before imaging the BEC.

\end{document}